# Ionospheric correction of space radar data

Mike Hapgood

Science & Technology Facilities Council, Rutherford Appleton Laboratory, Harwell Science and Innovation Campus, Didcot, Oxfordshire, OX11 0QX, United Kingdom

Abstract

Radar is a critical tool for maintaining knowledge of the many objects in low Earth orbit and thus for maintaining confidence that societies around the world are secure against a variety of space-based threats. It is therefore important to raise awareness that LEO objects are embedded in the envelope of relatively dense plasma that co-rotates with the Earth (ionosphere-plasmasphere system) and thus accurate tracking must correct for the group delay and refraction caused by that system. This paper seeks to promote that awareness by reviewing those effects and highlighting key issues: the need to customise correction to the altitude of the tracked object and prevailing space weather conditions, that ionospheric correction may be particularly important as an object approaches re-entry. The paper outlines research approaches that should lead to better techniques for ionospheric correction and shows how these might be pursued in the context of the EURIPOS initiative.

1.  **INTRODUCTION**

The accurate tracking of space objects close to the Earth is increasingly recognised as a function that is essential to the security of human activities – in particular, the security of space-based infrastructure that supports many activities vital to functioning of economies and societies around the world. These objects include active and obsolete spacecraft, inert items associated with launch and deployment of spacecraft as well as debris arising from failures, explosions and collisions. Radar tracking is a key technique for monitoring the orbits of all such objects in low Earth orbit (altitudes < 2000 km), providing information on the range and bearing of objects and on their line-of-sight velocity. As with any orbit determination technique, it is critical to

___________________________________________________

understand the accuracy of tracking data and to correct for systematic errors. Any uncertainty in orbit determination will lead to a time-growing error in the predicted position of the object. The radar signals used to track space objects are significantly perturbed as they propagate through the ionosphere-plasmasphere system. Thus it is important to correct for systematic errors that arise from that propagation and also to quantify any stochastic errors caused by that propagation.

We note that it is timely to improve our understanding of the ionospheric correction of space radar data. This has the potential to be a valuable input into the emerging programmes on space situational awareness (SSA) and, in particular, the European SSA programme that started in 2009 under the leadership of the European Space Agency. This programme seeks to develop European capabilities in both space surveillance and space weather. The initial requirements studies for that programme have demonstrated that ionospheric correction of space radar data is a topic that needs more attention. It has not been included in the requirements considered by previous ESA space weather studies, unlike all other aspects of space weather in the remit of the SSA programme (e.g. see the synthesis of user requirements in Hapgood (2008)). Thus it is timely to raise greater awareness of the capabilities that exist to correct for space weather effects on the tracking of space objects and of the potential for improved capabilities. There is major scope to advance this topic to the mutual benefit of the space weather and space surveillance communities.

To set the scene this paper will first outline the peculiar sensitivity of orbit predictions to uncertainties and systematic errors in tracking data. This is followed a mathematical review of the effects that perturb radar signals as they traverse the plasma which forms the ionosphere and plasmasphere. We show that these effects are very significant for the accuracy of tracking data and thus create the requirements for ionospheric correction of tracking data. We emphasise the need to customise such corrections to the altitude of a particular spacecraft. This proves to be a critical requirement which drives the need for improved understanding of the three-dimensional state of the ionosphere-plasmasphere system. We also note the need to support campaigns for critical events such as re-entries of major objects and explore the value of ionospheric corrections during those events. We outline the regular and irregular processes that control the state of the ionosphere-plasmasphere system and explore how our understanding of these processes may be improved through future research in the context of EURIPOS, a European initiative that seeks to promote an innovative approach to the study of the ionosphere-plasmasphere system.

Finally, please note that this paper has been written with two audiences in mind: one is the ionospheric community, while the other is the space surveillance community. With this in mind, the author has sought to provide a level of detail that will address the needs of both communities. Parts of the text therefore present a level of detail that goes beyond the need of one community or the other, but hopefully that enables both communities to understand the ideas presented here.

## 2. IMPACT OF TRACKING ERRORS ON ORBIT PREDICTIONS

Tracking errors are a key challenge in accurately predicting the future motion of space objects. Such predictions are extremely sensitive to small uncertainties in the velocity of the object as those uncertainties lead to a growing uncertainty in the future position of the object. For example an uncertainty of just 1 mm s$^{-1}$ in velocity ( $\leq 0.00002\%$ of the velocity of an object in low Earth orbit) will accumulate to an error of 86 metres after just one day and over 600 metres in a week.

However, the velocity of any space object is intimately related to its position with respect to the primary body that it orbits via Kepler's Laws. In particular, the Third Law states that the square of the orbit period is proportional to the cube of the semi-major-axis. We now express this as:

$$a^3 = \left(\frac{GM}{4\pi^2}\right)T^2 \qquad (1)$$

where T is the orbital period, a is the semi-major axis and GM is the gravitational coefficient of the primary body. The latter is just the product of the gravitational constant G and the primary body mass. We can re-arrange this in differential form thus:

$$3a^2\Delta a = 2\left(\frac{GM}{4\pi^2}\right)T\Delta T \qquad (2)$$

and dividing each side of equation 2 by the corresponding side of equation 1 we can obtain:

$$\frac{\Delta T}{T} = \frac{3}{2}\frac{\Delta a}{a} \qquad (3)$$

Thus an uncertainty in the distance of the object from the primary also implies an uncertainty in its orbital period and hence its velocity. If we consider the case of a circular orbit, which is a reasonable approximation for many objects in low Earth orbit, we can write the velocity as $V = 2\pi a/T$, where a is now the radius of the circular orbit. Thus we can estimate the uncertainty in velocity as:

$$\Delta V = 2\pi \left( \frac{\Delta a}{T} - \frac{a \Delta T}{T^2} \right) \quad (4)$$

Using equation 3 to substitute for ΔT/T we obtain

$$\Delta V = -\frac{\pi \Delta a}{T} \quad (5)$$

The negative sign arises because the velocity of an orbiting object increases as a decreases, but this can be ignored when estimating uncertainties. Thus if we assume a small uncertainty in Δa ~ 10 metres and a typical low Earth orbit period T ~ 90 minutes, we have ΔV ~ 6 mm s$^{-1}$, which would accumulate to a position error of 500 metres within a day. This demonstrates the need to make accurate measurements of the positions of space objects in order to make good predictions of their future motion.

## 3.  PLASMA EFFECTS ON SPACE RADARS

### 3.1 Outline of the problem

The basic physical concepts underlying the ionospheric correction of radar measurements are well outlined in publicly available material about spacecraft tracking (e.g. Air University, 2003). There are range errors arising from the group delay as signals pass through the plasma that forms ionosphere-plasmasphere system – the path delay familiar from GPS measurements, but potentially larger as many space radars operate at UHF frequencies. In addition, the signals experience significant refraction as they pass through the plasma surrounding the Earth. This leads to significant errors in the measured bearing and consequent errors in position (the signal does not follow a straight line) and velocity (the measured vector has wrong direction).

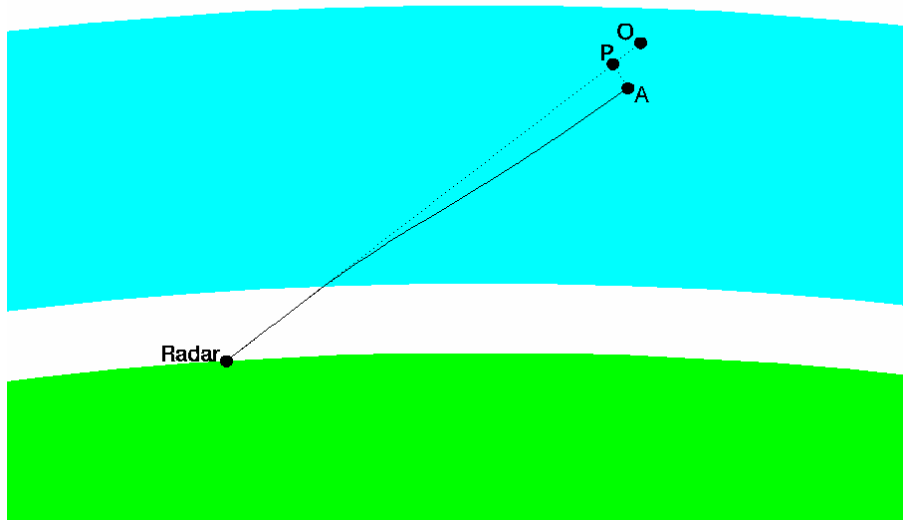

**Figure 1.** Schematic showing a ground based radar (left) tracking a space object in low Earth orbit (right). The radar signal (solid line) is refracted as it traverses the ionosphere (light blue). The distortion from a straight line is exaggerated (by a factor of 100) to provide a clear illustration of the effect.

These effects are outlined in Figure 1. The radar receives an echo from an object at a given range and direction. Thus the apparent position of the object (point O) can be obtained by projecting a straight line over the measured range and direction – as indicated by the dashed line in Figure 1. But the refraction causes the signal to follow a curved path – as indicated by the solid line in Figure 1. The actual position of the object (A) is closer to the radar because of group delay (line OP) and displaced sideways by refraction (line PA).

### 3.2 Group delay

The group velocity of a radar pulse travelling through a plasma is

$$V_g = c\left(1 + \frac{v_p^2}{v^2}\right)^{-1/2} \qquad (6)$$

where $v$ is the radar frequency and $v_p$ is the plasma frequency. Thus the one-way travel time of a radar pulse between the radar and the space object is

$$T = \int \frac{1}{V_g} dl = \frac{1}{c}\int \left(1 + \frac{v_p^2}{v^2}\right)^{1/2} dl \approx \frac{L}{c} + \frac{1}{2cv^2}\int v_p^2 dl \qquad (7)$$

where the integral over dl is taken over the range from the radar to the object. The first term (*L/c*) is the unperturbed travel time over range L, while the second term is the one-way group delay time. If we replace the plasma frequency by electron density using the standard formula $v_p = 8.978 n_e^{1/2}$, the group delay can be expressed as a distance thus

$$\Delta L = \frac{80.6}{2v^2}\int n_e dl \qquad (8)$$

where the integral is the total electron content along the one-way signal path and $\Delta L$ is equivalent to line OP in Figure 1. If we consider the case of a horizontally stratified ionosphere and neglect the curvature of the Earth, this can be expressed as

$$\Delta L = \frac{40.3\sec(z_0)}{v^2}\int_0^h n_e(x)dx \qquad (9)$$

where $z_0$ is the zenith distance of signal path and the integral is now taken over altitude x from the surface to the altitude of the object (h). Thus the group delay is related to the path-integrated electron content.

For a UHF radar observing a spacecraft in low Earth orbit $\Delta L$ will have a value ranging from a few metres to several hundred metres depending on the state of the ionosphere and altitude of the object. Examples of such radars include the Fylingdales radar in the UK (which operates around 420 MHz) and the radar proposed for the European SSA programme (435 MHz). This can increase to several kilometres for mid-latitude VHF radars such as the French Graves radar, which operates at 143 MHz. Indeed early work on this problem (Burns and Fremouw, 1970) suggested that group delay errors at VHF frequencies could be as large as 23 km at equatorial latitudes, where the largest path-integrated electron content may be found.

### 3.3 Refraction

The general principles of ionospheric refraction are well-outlined in the discussion of radio astrometric techniques by Green (1985) and the following outline draws heavily on the ideas in his book.

If we consider the case of a horizontally stratified ionosphere and neglect the curvature of the Earth, the transverse displacement of the radar signal, equivalent to line PA in Figure 1 is given by:

$$\Delta e(h) = \tan(z_0)\sec(z_0)\int_0^h (1 - n(x))dx \qquad (10)$$

where $\Delta e(h)$ is the transverse displacement at altitude h above the surface, $z_0$ is the zenith distance of the object as observed at the radar and $n(x)$ is the refractive index of the ionosphere at altitude x. $n(x)$ is taken as unity below the ionosphere; in the ionosphere, the plasma phase refractive index applies thus:

$$n(x) = \left(1 - \frac{v_p^2}{v^2}\right)^{1/2} \qquad (11)$$

where $v$ is the radar frequency and $v_p$ is the plasma frequency at altitude x. The latter is set by plasma number density $n_e$ at that altitude, i.e. $v_p = 8.978 n_e^{1/2}$, where all quantities are expressed in SI units.

The critical factor setting the order of magnitude of $\Delta e(h)$ is the integral $\int_0^h (1 - n(x))dx$. As with the group delay, this factor is related to the path-integrated electron content. This as follows using the plasma phase refractive index from equation 11 and the relationship $v_p = 8.978 n_e^{1/2}$.

$$\Delta p = \int_0^h \left(1 - \left(1 - \frac{v_p^2}{v^2}\right)^{1/2}\right)dx \approx \int_0^h \left(1 - \left(1 - \frac{v_p^2}{2v^2}\right)\right)dx = \frac{40.3}{v^2}\int_0^h n_e dl \qquad (12)$$

Substituting this into equation 10, and taking the integral over height rather than the path, we obtain:

$$\Delta e(h) = \frac{40.3}{v^2}\tan(z_0)\sec^2(z_0)\int_0^h n_e(x)dx \qquad (13)$$

Thus $\Delta e$ will have very similar values to group delay correction, $\Delta L$, discussed in the previous section, i.e. at UHF frequencies it will vary from a few metres to several hundred metres depending on the state of the ionosphere and the altitude of the object. It can increase to over a kilometre for radars working at VHF frequencies. However, there is one big difference. Refraction generates a systematic position error perpendicular to the line of sight to the object, whereas the group delay generates errors only along the line of sight.

### 3.4 Spacecraft embedding in the ionosphere

It is important to recognise that spacecraft in low-Earth orbit are embedded in the ionosphere-plasmasphere system. Thus radar signal propagation is affected only that part of the ionosphere and plasmasphere which lies between the radar and the spacecraft. This is reflected in the form of equations 9 and 13 – the integrals are carried out over the altitude range from the surface to the object. The ionospheric correction is a function of the object altitude. This is illustrated in Figure 2, which shows an estimate of the group delay correction $\Delta L$ using a recent daytime electron density profile (10:30 UTC, 25 August 2009) from the Chilton ionosonde in the UK (51.6° N, 1.3° W). The profile is extrapolated to 1000 km altitude using a simple exponential decay with increasing altitude above the peak density. The refraction correction $\Delta p$ is of similar size and exhibits the same variation with altitude and frequency.

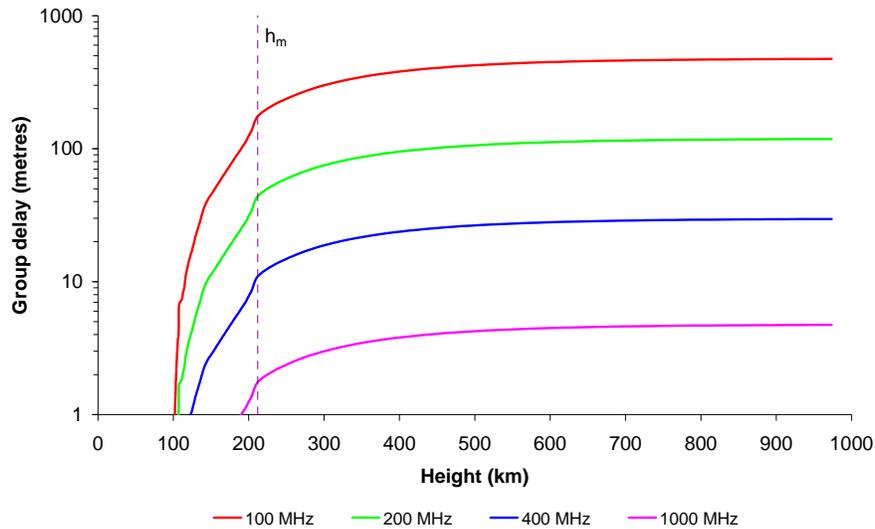

**Figure 2. Variation of group delay $\Delta L$ with altitude for four different frequencies under solar minimum daytime conditions. The vertical dashed line indicates $h_m$, the altitude of peak plasma number density.**

This figure is just one example, taken in the daytime under current (2009) solar minimum conditions. The level and form of the curves shown in Figure 2 will change with many factors:

    a. Local time. During the night-time we would expect the onset of significant ionospheric corrections to move to higher altitudes (~300

km rather than 200 km) and also for the size of the correction to reach lower peak values. This is the natural response to the normal daily changes in the ionosphere. At night-time the plasma number density drops in the absence of ionising radiation (extreme ultra-violet and X-rays) from the Sun and existing ionisation persists overnight only at higher altitudes where the recombination rates are low.

b. Seasonal changes. The seasonal changes in solar visibility and elevation naturally modulate the production of ionisation. However, the loss of ionisation is modulated by the complex seasonal changes in the thermosphere. Thus ionosphere exhibits a complex seasonal behaviour, and this will in turn, lead to a complex seasonal behaviour in the ionospheric corrections.

c. Solar cycle. We would expect a significant increase in the ionospheric corrections (perhaps up to factor 10) as ionospheric electron densities increase in response to high fluxes of extreme ultra-violet and X-rays from the Sun.

d. Solar flares. These intense bursts of X-rays from the Sun produce short-lived (10 minutes to an hour) enhancements of the ionosphere, especially at lower altitudes. They are likely to increase the ionospheric corrections.

e. Geomagnetic storms. These have profound effects on the ionosphere because they transform the properties of the thermosphere, which is the key factor controlling loss of ionisation. During storms auroral heating can reverse thermospheric winds so that they flow away from polar regions rather than the equator (where solar heating normally dominates). The auroral heating can also inject molecular species (e.g. NO and $O_2$) into the upper atmosphere (where atomic oxygen is normally the main species). These molecular species can significantly increase loss of ionisation through dissociative recombination. The ionosphere exhibits a complex response to these changes, sometimes exhibiting increased densities (especially in the early phase of a storm), but more often showing lower densities. Thus storms will have a complex effect on ionospheric corrections.

f. Solar energetic particle events. These generate extra ionisation in the lower ionosphere, especially in high-latitude regions where the Earth's magnetic field focuses the precipitation of these particles. In principle, this ionisation could change the ionospheric corrections. However, much of this ionisation is produced at altitudes below 100 km, so the resulting plasma will be collision-dominated and thus the usual plasma refractive index (equation 11) may not apply.

  g. Tropospheric coupling. There is growing evidence that tropospheric phenomena have a significant effect on the ionosphere, e.g. through the upward propagation of tidal modes and acoustic gravity waves (Immel et al, 2009) as well as possible changes in the altitudes of ionospheric layers as an upper atmosphere response to climate change (Laštovička et al, 2008). This is an expanding area of research that may have implications for ionospheric corrections.

The critical conclusion is that the ionospheric correction of radar measurements of spacecraft positions in low-earth orbit should be scaled to the altitude of the spacecraft. It is important to have some knowledge of the altitude distribution of ionisation and not just the electron content integrated over the whole altitude range of the ionosphere-plasmasphere system. The latter quantity is the total electron content (TEC) measured using GPS observations. The GPS spacecraft operate at very high altitude (around 20000 km altitude) so the ground-based observations of their signals detect the effect of the whole ionosphere-plasmasphere system. Only that part of the measured TEC which lies between the radar and the spacecraft is applicable to correction of tracking data.

Thus we conclude that the ionospheric correction of radar tracking of space objects will be greatly aided by a better understanding of the vertical distribution of ionisation in the ionosphere-plasmasphere system. Such understanding should include an objective of developing efficient algorithms for the partition of TEC at any altitude, i.e. determining what fractions of total TEC lie below and above any altitude. Future work in these areas should include a comprehensive study on how this partition will change in response to many factors that drive ionospheric variations as discussed above. It is particularly important to understand how correction algorithms should be adapted during major space weather activity, e.g. in response to solar flares, solar energetic particle events and geomagnetic storms.

### 3.5 Plasmaspheric effects

It has been known for some time (Lunt et al, 1999) that the plasmasphere can sometimes contribute a significant fraction of the TEC measured by GPS techniques. This arises because the line of sight from the receiver to a GPS spacecraft can include a long path through the plasmasphere. Since the plasmasphere is structured by the geomagnetic field, this path length depends on the field topology between the receiver and the spacecraft; thus it varies with the location of the receiver and the spacecraft. The relative TEC contributions from ionosphere and plasmasphere also respond to the usual factors of local time, season and solar cycle. The plasmaspheric contribution is of greatest significance when the ionosphere is at its weakest, e.g. at night and at solar minimum conditions. In such cases, the plasmasphere can contribute up to 50% of TEC (Lunt et al, 1999). Fortunately, these are also the cases where the ionospheric correction will be at its smallest. Nonetheless it is important to be able to identify such cases and maintain the accuracy of tracking data by applying a suitably small or even zero correction.

The plasmasphere also exhibits variations in response to space weather events such as geomagnetic storms. It can be severely eroded during strong geomagnetic storms and gradually recovers after the storm, but the process of recovery is not yet well-understood. Thus it is important to gain a better understanding of the storm-time erosion and recovery of the plasmasphere, so that the impact of these phenomena on ionospheric corrections can be better modelled.

The plasmasphere is magnetically coupled to the topside ionosphere (i.e. the region above peak plasma number density) such that there is a gradual transition between the two regions. Furthermore the topside ionosphere is a source for the material that fills the plasmasphere. Thus our understanding of the plasmasphere is closely linked with the understanding of the topside. Furthermore, the topside is major contributor to TEC and thus to the ionospheric correction of tracking data. This is well-illustrated in Figure 2, which shows that most of ionospheric correction accumulates above the altitude of peak plasma number density ($h_m$). This large contribution arises because the scale height for changes in number density is much larger in the topside ionosphere than below $h_m$. There is much current interest in developing better models of the topside (e.g. see the papers by Belehaki et al (2009) and Kutiev et al (2009) in this issue) and this could be important in developing a better understanding of how to partition TEC to derive accurate ionospheric corrections for tracking data. We note in particular that Kutiev et al (2009) suggest that current methods can significantly underestimate the topside scale height, perhaps by factor 3. A larger scale height would significantly change the correction curves shown in Figure 2. It would create a stronger requirement to partition TEC to the altitude of the space object being tracked and make to ionospheric corrections based only that part of TEC accumulated below the object.

### 4.  POSSIBLE APPROACHES

It is essential that any new approach to ionospheric corrections can be transferred to the operational domain of space surveillance. We must look for solutions that will work well under realistic constraints on time, processing power and data availability. It is important that corrections are available in near-real-time so that they can be applied promptly to new tracking data and thus those data can be used to predict the short-term (say one day ahead) motion of space objects. This would be very important in some cases, for example as a major object in low Earth orbit approaches re-entry. In such cases frequent updates to orbit predictions will be important in estimating the regions at risk from impact following re-entry. We note, in passing, that the re-entry case is particularly challenging for ionospheric corrections, since the object will then be at low altitudes such that partition of TEC will be a very significant factor in the corrections.

It is also important to know the quality of any ionospheric corrections applied to tracking data. The uncertainties on ionospheric corrections contribute to the overall error budget of tracking and thus need to be included in the estimation of the uncertainties in the predicted motion of space bodies. There will, inevitably, be occasions when the uncertainties in ionospheric corrections become large, e.g. due to unavailability of ionospheric sensors or due to space weather perturbations of the ionosphere. Thus it is essential to have good methods for assessing these uncertainties.

Thus future work on ionospheric correction products should explore how our improving knowledge of the ionosphere-plasmasphere system can be exploited to generate near-real-time corrections and give an estimate of their accuracy. An obvious and important issue is to minimise use of iterative procedures and to seek products that facilitate a simple mapping of observed spacecraft positions to a corrected position. It is important to note that such corrections change the estimated altitude of the spacecraft, but only by a small amount (typically <1km). Thus any partition of TEC calculated using the apparent altitude should remain valid; there is no need to iterate that partition to take account of the corrected altitude. Thus the key aim of research should be to develop an algorithm that can easily map between observed and corrected spacecraft position. It may be useful to simulate the tracking processes, e.g. taking various models of the ionosphere-plasmasphere system and tracing radar signal propagation to the observed and actual spacecraft locations. Such simulations may also prove useful as a tool for assessing the uncertainties in ionospheric corrections.

As we noted above, it is important that a new techniques for ionospheric correction are robust against missing data. It is possible to obtain a high level of real-time availability (say >95%) for data from ionospheric sensors such as ionosondes and GPS receivers. But 100% availability is not realistic; there will always be some data loss due to problems with instruments, power sup-plies and data links. Ionosondes will also lose data on occasions when there are high levels of radio wave absorption in the middle atmosphere (D-region); such absorption can occur at mid-latitudes during strong solar flares and also when there is precipitation of high energy electrons from the radiation belts. GPS receivers are less affected by natural interference but will lose data when a fault on a GPS spacecraft leads to errors in phase measurements; once detected by the GPS ground segment, these faults are indicated by a flag in the data stream from the affected spacecraft and thus the bad data can be ignored. However, in the worst case, this can take several hours.

The imperfect nature of the data from real-time sensors is an important issue in developing methods for ionospheric correction. The underlying ionospheric model should use all the available near-real-time data in order to obtain the best knowledge of the current state of the ionosphere. But that knowledge should degrade gracefully as the amount of available data is reduced, i.e. the model should continue to provide valid and consistent information while allowing the uncertainties to increase. This may be done by a mixture of methods. One important element is to have a degree of redundancy in the sensor network, so that missing data can be replaced by interpolation between adjacent sensors. The corrections should not be critically dependent on any single sensor. Another important element, which will become important if too many sensors are lost, is to use a climatological model to estimate what the sensors would have seen given their location and the time of the observation. The transition from use of real-time data to a climatological model should be done gradually by an interpolation from the last measured data to values predicted by the model.

Future research should also consider whether the ionospheric correction should be directly embedded in the modelling of the orbits of space bodies, rather than being a correction applied before tracking data are input to orbits models. Orbit models typically follow an assimilative approach in which the underlying physical model is constantly refined and propagated forward in time as new data are received. It may therefore be possible to include the ionospheric correction in an assimilative model, so that external data on ionospheric conditions is an input alongside raw tracking data. The model could then estimate the ionospheric correction internally and provide estimates of the correction, and its uncertainty, as an output. If this approach proves feasible, it would be worth assessing if these outputs could provide useful feedback on ionospheric conditions along the track of the space body.

**5. DISCUSSION AND SUMMARY**

The analysis presented above shows that it is important to understand the ionospheric corrections that should be applied to the apparent positions of space objects tracked by radar. The radar echo received back from any space object is subject to both group delay effects and to refraction in the ionosphere. Both effects generate systematic errors in the measured range, whilst refraction can also generate systematic position errors perpendicular to the line of sight. These errors will also be systematic along the track of the object since the ionosphere usually exhibits only a slow variation in latitude and longitude. At UHF frequencies the ionospheric correction can vary from a few metres to several hundred metres depending on the altitude of the object and the state of the ionosphere. The correction will be even bigger at VHF frequencies possibly ranging up to several kilometres.

The ionospheric corrections are dependent on the electron content that lies on the path between the radar and the space object. However, the object will be embedded in the ionosphere-plasmasphere system. Thus to obtain an accurate correction, it is essential to identify that part of the total electron content that lies between the radar and the space object. To determine this partition of TEC we need to have a good understanding of the distribution of plasma in the ionosphere and plasmasphere. That understanding is already a key objective of the EURIPOS initiative and so this is a topic to which EURIPOS can make important contributions. These research outputs are likely to be a valuable input to the plans to develop a European capability for space surveillance, including radar tracking.

A key target for this future research will be to improve our understanding how the distribution of plasma in the ionosphere and plasmasphere responds to the complex set of variations that control the state of the ionosphere. These include (a) regular cycles on diurnal, season and solar cycle timescales and (b) irregular variations due to several space weather phenomena. It is important to study how space weather phenomena affect the ionospheric correction. It has long been reported that geomagnetic storms can seriously degrade maintain knowledge of the positions of space objects. For example, it was widely reported that, during the great magnetic storm of 13/14 March 1989, the US tracking agency (then known as NORAD) lost knowledge of the motion of over 1600 objects. This is usually attributed to an inability to model changes in the mass density of the thermosphere (and hence spacecraft drag) during the storm. However, given the sizes of the ionospheric corrections discussed above, it is appropriate to speculate whether uncertainties in ionospheric corrections could also contribute to loss of knowledge about the motion of space objects. This is also a potential issue for further research that can be addressed within the EURIPOS initiative.

An emerging issue in ionospheric science is the idea that there are ionospheric variations driven by activity in the underlying troposphere. This is an exciting topic of current research and should be monitored in case it proves to affect the accuracy of ionospheric corrections. We should also consider the possible impact of climate change on ionospheric corrections. It has long been suggested that heating of the lower atmosphere by greenhouse gases will lead to a cooling, and thus a lowering, of the upper atmosphere and there is now growing evidence that this is happening. Thus it is important to assess how this will affect ionospheric corrections, e.g. will it simply shift the curves shown in Figure 2 to lower altitudes or will it change their shape. We note in passing that climate change in the upper atmosphere will also affect atmospheric drag; this reinforces the importance of studying how climate change in the upper atmosphere will impact plans for space situational awareness activities. Thus there is considerable scope for EURIPOS to work with the SSA community to mutual benefit of both.

We conclude that the ionospheric correction of radar tracking of space objects is an important application for which EURIPOS has the potential to provide improved techniques. This research should include objectives that recognise the operational needs of the space surveillance community. In particular, it should address (a) the development of operationally effective and robust techniques for estimating the ionospheric correction using data on the three-dimensional structure of the ionosphere, (b) provision of an estimate of the uncertainty in any ionospheric correction and (c) the integration of these techniques into assimilative models of space orbits.

We also note that there are special cases where high accuracy ionospheric corrections are needed. An obvious example is during the re-entry of a major object in low Earth orbit. Accurate orbit predictions are then needed to manage impact risk and thus create a requirement for high quality ionospheric corrections. This will be reinforced by the low altitude of the object as that implies that it will be very important to partition TEC measurements to the object altitude. In such cases, these correction techniques should be capable of supporting campaign operations, e.g. ingestion of real-time data from additional ionospheric sensors, use of additional resources to generate and disseminate high accuracy products.

## 6. ACKNOWLEDGEMENTS

I thank Anna Belehaki and Lili Cander for encouraging me to contribute this paper to the EURIPOS special issue. I also thank the many colleagues involved in the ESA and US space situational awareness programmes who have provided me with the inspiration to explore this fascinating subject. I also wish to acknowledge Bill Mish (ex-NASA Goddard Space Flight Center) who many years ago introduced me to the importance of assimilative techniques in orbit modelling.

R e f e r e n c e s

Air University (2003), *Air University Space Primer, Chapter 6: Space Environment,* http://www.au.af.mil/au/awc/space/primer/space_environment.pdf

Belehaki, A., I. Kutiev, B. Reinisch, N. Jakowski, P. Marinov, I. Galkin, C. Mayer, I. Tsagouri and T. Herekakis (2009), *Verification of the TSMP-assisted Digisonde (TaD) topside profiling technique*, Acta Geophys., this issue.

Burns, A.A. and E.J. Fremouw (1970), *A real-time correction technique for transionospheric ranging error*, IEEE Trans. Antennas Propag., AP-18, 785-790.

Green, R.M. (1985), *Spherical astronomy*, Cambridge University Press, Cambridge, UK. DOI 10.2277/0521317797

Hapgood, M. (2008), *Nano Satellite Beacons for Space Weather Monitoring: Final Report*, Report of ESA contract 18474/04/NL/LvH. http://epubs.stfc.ac.uk/work-details?w=42988

Immel, T. J., S. B. Mende, S. L. England, P. M. Kintner, M. E. Hagan (2009), *Evidence of Tropospheric Effects on the Ionosphere*, Eos Trans. AGU, 90(9), 69, DOI 10.1029/2009EO090001.

Kutiev, I, P. Marinov, A. Belehaki, N. Jakowski, B. Reinisch, C. Mayer and I. Tsagouri, *Plasmaspheric electron density reconstruction based on the Topside Sounder Model Profiler*, Acta Geophys., this issue.

Laštovička, J., R.A. Akmaev, G. Beig, J. Bremer, J.T. Emmert, C. Jacobi, M.J. Jarvis, G. Nedoluha, Y.I. Portnyagin, T. Ulich (2008), *Emerging pattern of global change in the upper atmosphere and ionosphere*, Ann. Geophys. 26, 1255-1268. www.ann-geophys.net/26/1255/2008/

Lunt, N, L. Kersley and G.J. Bailey (1999), *The influence of the protonosphere on GPS observations: Model simulations*. Radio Sci. 34, 725-732, DOI 10.1029/1999RS900002.